\documentclass[aps,letterpaper,twocolumn,pra,footinbib,floatfix,showpacs]{revtex4}
\usepackage{amsmath}
\usepackage{amsfonts}
\usepackage{amssymb}
\usepackage[scanall]{psfrag}
\usepackage{psfrag}
\usepackage{graphicx}
\usepackage{color}
\usepackage[caption=false]{subfig}
\usepackage{bbold}
\usepackage{prettyref}

\usepackage{yypreamble}

\newcommand{\oUt}{{\scriptstyle\p{U t}}}

\begin{document}

\title{Heating from Continuous Number Density Measurements in Optical Lattices}

\author{Yariv Yanay}
\affiliation{Laboratory of Atomic and Solid State Physics, Cornell University, Ithaca NY 14850}
\author{Erich J. Mueller}
\affiliation{Laboratory of Atomic and Solid State Physics, Cornell University, Ithaca NY 14850}

\date{\today}

\begin{abstract}
We explore the effects of continuous number density measurement on atoms in an optical lattice. By integrating a master equation for quantum observables, we calculate how single particle correlations decay. We consider weakly- and strongly- interacting bosons and noninteracting fermions. Even in the Mott regime, such measurements destroy correlations and increase the average energy, as long as some hopping is allowed. We explore the role of spatial resolution, and find that the heating rate is proportional to the amount of information gained from such measurements. 
\end{abstract}


\maketitle

\section{Introduction}

One of the most important recent advances in cold atom experiments is single-site resolved imaging in optical lattices \cite{Nelson2007,Gericke2008,Bakr2009,Sherson2010}. Presently these techniques are destructive, and do not directly yield dynamical information. While back-action from measurement is inherent to quantum mechanics, a less destructive local probe is desirable, as it would enable whole classes of new experiments \cite{Patil2014}. Here we explore the ultimate limits on such a program, calculating how correlations evolve during ideal continuous local density measurements. We quantify the heating in weakly and strongly interacting gases.

Quantum back action arises when the system's energy eigenstates and the measurement operator do not commute. While this back action can be a useful resource \cite{Naik2006,Barontini2013,Fowler2012,Itano1990,Fischer2001,Schafer2014,Dalvit2002}, more often it leads to unwanted heating or decoherence \cite{Giovannetti2004,Hatridge2013,Gambetta2008,Mendoza-Arenas2013,Dalvit2002a}. We consider measuring the local density of atoms in a lattice. Such a measurement localizes individual atoms to single sites, projecting their wavefunctions to superpositions of momentum states. As noted by Poletti et al., \cite{Poletti2012}, in the long-time limit, this results in an infinite temperature system where all kinetically accessible many-body Fock states are equally likely.

We quantify the approach to this steady state using a master equation for the non-unitary evolution of the density matrix and observables. In the weakly-interacting limit, where atoms are highly delocalized, off-diagonal elements of the single particle correlation function fall off exponentially with time. In the strongly-interacting limit, where number density is nearly a good quantum number, we find slower evolution: an exponential stage where quasiparticle momenta are scrambled is followed by a slow proliferation of excitations and a parallel decay in correlations.

This heating arises even if the measurement photons are never detected. Thus our formalism is nearly identical to that used by others \cite{Pichler2010,Poletti2013,Poletti2012,Pichler2013,Cai2013,Schachenmayer2014,Diehl2008,Mekhov2012} to study spontaneous off-resonant light scattering in an optical lattice. Other works approached the subject using different formalisms \cite{Gerbier2010,IvanaVidanovic2014,Knap2014,Riou2013,Barmettler2011}. 

Our principal results come from from applying variants of the Bogoliubov approximation and calculating the time dependence of single particle correlation functions. Such approaches work well in both the weakly and strongly interacting limits, but do not accurately describe intermediate coupling strength \cite{Hohenberg1965}. Previous works used one-dimensional numerical techniques or assumed slow photon scattering rates. Our approximations apply to three-dimensional systems and do not restrict the scattering rate. Our results are consistent with previous studies, and in many places extend our understanding. For example, the doublon-holon picture we present in \pref{sec:strong} gives a clear explanation of the two timescale that have been previously observed in the Mott regime \cite{Poletti2012} and allows us to quantify the decay rates associated with each.

Our paper is organized as follows. In \pref{sec:model}, we introduce our model and the master equations used to calculate the evolution of the system. From the form of the expressions we make some general observation about the evolution of momentum states and single-particle correlations. In \pref{sec:weak}, we use a Bogoliubov approach to integrate the master equations for weakly interacting bosons. In \pref{sec:strong} we extend these calculations to the Mott regime through a doublon-holon formalism. Finally, in \pref{sec:lowres} we consider the use of longer-wavelength light in measurement, exploring the trade-off between information extracted from the system and the heating caused by measurement. 

\section{Model\label{sec:model}}

We model the optical lattice system with the single-band Hubbard model,
\begin{equation} \begin{split}
\hat H & = -J \sum_{\avg{i,j}} \p{ \hat a_{i}\dg \hat a_{j} + \hat a_{j}\dg \hat a_{i}} \\& \qquad + \sum_{i}\tfrac{U}{2}\hat n_{i}\p{\hat n_{i} - 1} - \p{\mu - 2J D}\hat n_{i}
\\ & = \sum_{k} \p{J \gep_{k} - \mu}\hat n_{k} + \tfrac{U}{2}\sum_{i}\hat n_{i}\p{\hat n_{i} - 1}
\end{split} \end{equation}
where $\hat a_{i}$ ($\hat a_{i}\dg$) is the annihilation (creation) operator at site $i$; $\avg{i,j}$ are nearest neighbor sites $i$ and $j$; $\hat n_{i} = \hat a_{i}\dg \hat a_{i}$ is the occupation operator at site $i$; $\hat n_{k} = \hat a_{k}\dg \hat a_{k}$ is the occupation of the momentum mode $\vk$; and $2D$ is the number of nearest neighbors per site. $J$, $U$ and $\mu$ are the hopping energy, interaction energy and chemical potential, respectively. Here we define $\hat a_{k} = \frac{1}{\sqrt{N_{s}}}\sum_{i}e^{i \vk\cdot \vec r_{i}} \hat a_{i}$, summing over $N_{s}$ sites at positions $\vec r_{i}$. The kinetic energy is given by $J \gep_{k} = J\sum 4\sin^{2}\p{\vk\cdot\Delta\vec r/2}$ where the sum is over all lattice basis vectors $\Delta \vec r$.

We model the measurement process as an additional term of the form $ \hat H_{I} = \gl\sum_{\ga}\p{\hat c_{\ga} + \hat c_{\ga}\dg}\hat M_{\ga}$ where $\hat c_{\ga}$ are annihilation operators for a set of independent zero-temperature photon baths. For single-site resolved position measurements, we take $\hat M_{\ga} = \hat n_{i}$. We consider a more general operator in Section \ref{sec:lowres}. Following Gardiner \cite{Gardiner} we adiabatically eliminate the density matrix of the photons to derive a master equation for $\hat \rho$, the density matrix of the atoms,
\begin{equation} \begin{split}
\frac{d}{dt}\hat \rho = i \br{\hat \rho, \hat H} - \half \gamma\sum_{i}\br{\hat n_{i}, \br{\hat n_{i}, \hat \rho}},
\end{split} \end{equation}
where we have used that $\gl$ is real and $\hat n_{i}$ is Hermitian. Here $\gamma$ is an energy scale related to the measurement rate. It is proportional to $\gl$ and the density of photon states. A more detailed derivation is found in \cite{Pichler2010}.

While the density matrix contains all information about the system, it has an exponentially large number of terms. Thus it is more convenient to work with observables such $\hat n_{i}$, $\hat n_{i}^{2}$ that are experimentally accessible. Using $\avg{\hat O} = \Tr\br{ \hat \rho \hat O}$, the evolution of observables is governed by
\begin{equation} \begin{split}
\label{eq:dObs}
\frac{d}{dt}&\avg{\hat O} =   i \avg{\br{\hat H, \hat O}} - \half \gamma\sum_{i}\avg{\br{\hat n_{i}, \br{\hat n_{i}, \hat O}}}.
\end{split} \end{equation}

Most of our results concern bosonic atoms, though we briefly address the case of noninteracting, spinless fermions. Much of the intuition gained carries over to interacting fermions. Irregardless of statistics, each photon scattered localizes a particle, generically heating the system by increasing the kinetic energy.

Throughout, we assume a homogenous system.

\subsection*{Equations of motion for single-particle observables}

The single particle correlations can be studied in  momentum space or position space. In a homogenous system, the relevant observables evolve as
\begin{equation} \begin{split}
\label{eq:dnk}
& \frac{d}{dt}\avg{\hat n_{k}} = 
\\ & -2U\tfrac{1}{N_{s}} \sum_{p,q}\Im\br{\avg{\hat a_{p-q}\dg \hat a_{k+q}\dg \hat a_{p} \hat a_{k}}}
 - \gamma\p{\avg{\hat n_{p}} - \rho},
\end{split} \end{equation}
\begin{equation} \begin{split}
\label{eq:daiaj}
& \frac{d}{dt}\avg{\hat a_{i}\dg \hat a_{j}}  = \\ &  i U\p{\avg{\hat a_{i}\dg\hat n_{i} \hat a_{j}} - \avg{\hat a_{i}\dg \hat n_{j} \hat a_{j}}}
 - \gamma\p{\avg{\hat a_{i}\dg \hat a_{j}} - \rho \gd_{i,j}}.
\end{split} \end{equation}
where $\rho = N_{p}/N_{s}$ is the average occupation per site. These are related by $\avg{\hat a_{i}\dg \hat a_{j}} = \tfrac{1}{N_{s}}\sum_{k}e^{i\vk\p{\vec r_{i} - \vec r_{j}}}\avg{\hat n_{k}}$. Setting $i=j$ in Eq. \pref{eq:daiaj} produces the intuitively obvious result that the average density $\rho = \avg{\hat a_{i}\dg\hat a_{i}}$ is constant.

\subsection*{Energy gain}
Applying Eq.~\pref{eq:dObs} to the Hamiltonian, we find irrespective of interactions
\begin{equation} \begin{split}
\label{eq:dE}
\frac{d}{dt}\avg{E} = \frac{d}{dt}\avg{\hat H} & = \gamma  J \sum_{\vk} \gep_{k}\p{\rho - \avg{\hat n_{k}}}.
\end{split} \end{equation}
The instantaneous rate of energy gain depends only on the kinetic energy in the system. It is proportional to  the difference between the kinetic energy and the ``infinite-temperature'' kinetic energy of a system with $\avg{\hat n_{k}} = \rho$. 

\begin{figure}[htbp] 
   \centering
   \includegraphics[width=\columnwidth]{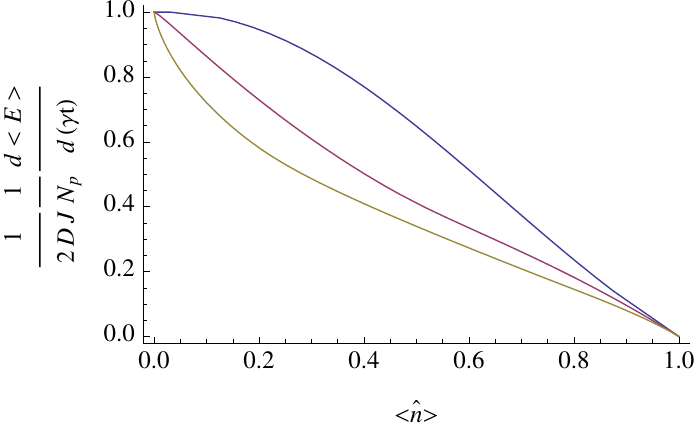} 
   \caption{(Color online) Initial rate of energy gain, $\frac{1}{2\gamma J D}  \frac{1}{N_{p}}\frac{d}{dt}\avg{E}$ as function of the filling fraction $\avg{\hat n}$ for spinless fermions. From top to bottom (blue, magenta, yellow), the rate for a  one-, two- and three-dimensional system. For noninteracting bosons, the initial rate is always $ \frac{1}{N_{p}}\frac{d}{dt}\avg{E} = 2\gamma J D$. Here $J$ is the hopping energy, $\gamma$ the measurement rate and $D$ is the dimension of the system.}
   \label{fig:dEdt}
\end{figure}

Eq. \pref{eq:dE} applies to both bosons and fermions. Fermions tend to have broader equilibrium momentum distributions, hence lower rates of energy gains. For free bosons at zero temperature $\avg{\hat n_{k}} = \gd_{k,0} N_{p}$, and one finds initially $\tfrac{1}{N_{p}}\frac{d}{dt}\avg{E} = 2\gamma J \times D  $. The equivalent result for free fermions is shown in Fig. \ref{fig:dEdt} as a function of filling. As $\avg{n_{i}}\to 0$, the fermionic rate approaches the bosonic rate.

This result differs from Eq.~(31) in \cite{Pichler2010}. There the off-resonant light scattering from the lattice can drive atoms to high bands, while we consider measurements that are engineered to keep atoms in the lowest band. For example, in \cite{Patil2014}, Raman side-band cooling rapidly returns atoms to the lowest band.

\subsection*{Non-interacting particles}
If $U = 0$, Eqs. \pref{eq:dnk} and \pref{eq:daiaj} are readily integrated,
\begin{equation} \begin{split}
\label{eq:dnNI}
\avg{\hat n_{k}} & = \p{\avg{\hat n_{k}}_{t=0} - \rho}e^{-\gamma t} + \rho
\\ \avg{\hat a_{i}\dg\hat a_{j}} & = \p{\avg{\hat a_{i}\dg\hat a_{j}}_{t=0} - \gd_{ij}\rho}e^{-\gamma t} + \gd_{ij}\rho.
\end{split} \end{equation}
These expressions hold for both noninteracting bosons and fermions, the only difference being initial conditions. The correlations decay exponentially with a time constant $\tau_{m} = 1/\gamma$ set by the measurement rate. The occupation of momentum states approaches a uniform distribution.

\section{Weakly Interacting Bosons\label{sec:weak}}

We extend our analysis to the weakly interacting case by a variant of the Hartree-Fock-Bogoliubov-Popov (HFBP) approach \cite{Griffin1996}. This approximation is well validated for static quantities in dimensions greater than one. It is a gapless model which includes interactions between atoms and discards some of the coherences between non-condensed particles.

Within this formalism we calculate $\avg{\hat n_{k}\p{t}}$ for $k\ne 0$, then infer the condensate density via $\rho_{c} = \tfrac{\avg{\hat n_{0}}}{N_{s}} = \rho - \tfrac{1}{N_{s}}\sum_{\vk\ne 0}\avg{\hat n_{k}}$. The occupation numbers evolve with Eq.~\pref{eq:dObs}, where we approximate $\hat H$ by the HFBP Hamiltonian
\begin{equation} \begin{split}
\label{eq:HBog}
& \hat H_{\rm HFBP}  =-\half[U]\p{2\rho - \rho_{c}}\avg{\hat n_{0}} 
\\ &  + \sum_{\vk\ne 0}\p{J\gep_{k} + U\rho_{c}}\hat n_{k} + \half U\rho_{c}\p{\hat a_{k}\hat a_{-k} + \hat a_{k}\dg\hat a_{-k}\dg}.
\end{split} \end{equation}
Evaluationg the commutators in Eq.~\pref{eq:dObs} yields
\begin{equation} \begin{split}
\frac{d}{dt} \avg{\hat n_{k}} &=   -2 U\rho_{c}\Im\br{\avg{\hat a_{k}\hat a_{-k}}} - \gamma\p{\avg{\hat n_{k}} - \rho}
\\ \frac{d}{dt} \avg{\hat a_{k}\hat a_{-k}} & = -2i\p{J\gep_{k} + U\rho_{c}} \avg{\hat a_{k}\hat a_{-k}}
\\ & \quad -iU\rho_{c}\p{\avg{\hat n_{k}} + \avg{\hat n_{-k}} + 1}
\\ & \quad  - \gamma\mat{\avg{\hat a_{k}\hat a_{-k}} + \tfrac{1}{N_{s}}\sum_{p}\avg{\hat a_{p}\hat a_{-p}}},
\end{split} \end{equation}
whereby the equations of motion of $\avg{\hat n_{k}}$ are coupled to those of $\avg{\hat a_{k}\hat a_{-k}}$.

Consistent with the Popov approximation, we replace $ \sum_{p}\avg{\hat a_{p}\hat a_{-p}} \to \rho_{c}$. This approximation only discards terms which vanish as $N_{s}\to \infty$. 

\begin{figure}[htbp] 
   \centering
   \includegraphics[width=\columnwidth]{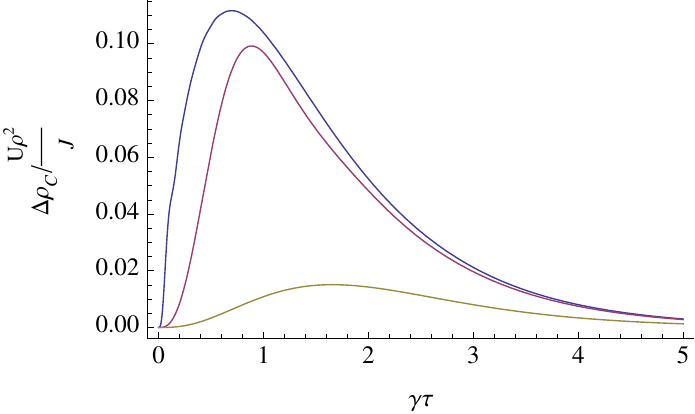} 
   \caption{(Color online) Corrections to the exponential decay of the condensate density, $\Delta\rho_{c} = \rho_{c}\p{t} - \rho e^{-\gamma t}$, induced by weak interactions. From top to bottom (blue, magenta, yellow) the corrections for $\gamma/J = 0.1, 1, 5$ in a three-dimensional cubic lattice. }
   \label{fig:WeakRho}
\end{figure}

These coupled equations can be perturbatively integrated for $U\ll J\gep_{k}$, yielding to first order in $U/J$,
\begin{equation} \begin{split}
\label{eq:dnkweak}
\avg{\hat n_{k}}& \approx \p{\avg{\hat n_{k}}_{t=0} - \rho}e^{-\gamma t} + \rho
\\ - & \frac{U\rho^{2}}{J}\frac{4 J^{2} \gep_{k}}{\gamma^{2} + 4J^{2}{\gep_{k}}^{2}}e^{-\gamma t}\p{1 - e^{-\gamma t}}
\\ + & \frac{U\rho^{2}}{J}{\frac{2\gamma\p{\gamma\sin^{2}\p{J\gep_{k}t}+J\gep_{k}\sin\p{2J\gep_{k}t}}}{\gep_{k}\p{\gamma^{2}+4J^{2}{\gep_{k}}^{2}}}}e^{-2\gamma t}.
\end{split} \end{equation}
By integrating over all momenta we find the condensate density. The leading behavior coincides with Eq.~\pref{eq:dnNI}. The deviation from this form is shown for a range of $\gamma/J$ in Fig.~\ref{fig:WeakRho}.
In three dimensions, this deviation is capped at $\rho_{c} - \rho e^{-\gamma t}\sim 0.1 \tfrac{U\rho^{2}}{J}$. Thus we expect detecting it would be very difficult.

\section{Strongly-Interacting Bosons\label{sec:strong}}

The low-energy states of the $U/J\gg 1$ Bose-Hubbard model with near integer filling, $\abs{\rho - \bar n} \ll 1$ for some integer $\bar n$, can be described by the subspace made up of 
states where the single site occupations are $\bar n, \bar n \pm 1$ \cite{Barmettler2012}. We model this behavior by introducing ``doublons'' and ``holons'' as hard-core particles representing an occupation of one-higher or one lower than the mean $\bar n$,
\begin{equation} \begin{split}
\hat a_{i} \to \sqrt{\bar n + 1}&\hat d_{i} + \sqrt{\bar n}\hat h_{i}\dg
\end{split} \end{equation}
with $\hat d_{i}^{2} = \hat h_{i}^{2} = \hat h_{i}\hat d_{i} = 0$. The names ``doublons'' and ``holons'' are motivated by the most common case, $\bar n = 1$. The effective Hamiltonian becomes
\begin{equation} \begin{split}
 \hat H_{\rm DH}  =  \sum_{k}&\br{ \half[U] + J \mat{\sqrt{\tilde n^{2} +\tfrac{1}{4}} + \half}\gve_{k}}\hat d_{k}\dg\hat d_{k} 
\\ & + \br{ \half[U] + J \mat{\sqrt{\tilde n^{2} +\tfrac{1}{4}} - \half}\gve_{k}}\hat h_{k}\dg \hat h_{k}
\\ &  + J \tilde n\gve_{k}\p{\hat d_{k}\hat h_{-k} + \hat h_{-k}\dg \hat d_{k}\dg}
\label{eq:HDH}
\end{split} \end{equation}
where $\hat d_{k}, \hat h_{k}$ are related to $\hat d_{i}, \hat h_{i}$ in the same way as $\hat a_{k}$ is to $\hat a_{i}$. Here $J\gve_{k} =J\p{\gep_{k} - 2D} = -2J\sum_{\Delta \vec r}\cos\p{\vk\cdot\Delta\vec r}$ is the kinetic energy and $\tilde n = \sqrt{\bar n\p{\bar n + 1}}$.

This structure is similar to that in Eq.~\pref{eq:HBog} with two exceptions. First, the Hamiltonian of Eq.~\pref{eq:HDH} allows for the creation of doublons and holons in pairs. Second, the hard-core constraints give non-bosonic commutation relations (see Eq.~\pref{eq:com} in the appendix). Neglecting non-coherent summations, these relations become
\begin{equation} \begin{split}
\br{\hat d_{k}, \hat d_{q}\dg} \to \gd_{k,q}\p{1 - 2\hat n^{d} - \hat n^{h}},
\label{eq:comapx}
\end{split} \end{equation}
where $\hat n^{d}~=~\tfrac{1}{N_{s}}\sum_{k}\hat d_{k}\dg\hat d_{k}$ is  the density of doublons and $\hat n^{h}$ the density of holons. This approximation is equivalent to a mean field theory of the interactions.

We apply Eq.~\pref{eq:dObs} to the Hamiltonian of Eq.~\pref{eq:HDH}, using the approximate commutation relations of Eq.~\pref{eq:comapx}. We decouple the equations for two-point functions from higher order correlations by assuming
\begin{equation} \begin{split}
\avg{\hat n^{d}\hat d_{k}\dg\hat d_{k}} \to n^{d}\avg{\hat d_{k}\dg\hat d_{k}}
\end{split} \end{equation}
and similarly for all combinations of $\hat n^{d}, \hat n^{h}$ with $\hat d_{k}\dg\hat d_{k}, \hat h_{k}\dg\hat h_{k}$ or $\hat d_{k}\hat h_{-k}$. Here $n^{d} = \avg{\hat n^{d}}$.

Under these assumptions, we find a set of coupled equations for $\avg{\hat d_{k}\dg \hat d_{k}}$, $ \avg{\hat h_{k}\dg \hat h_{k}}$, $\avg{\hat d_{k} \hat h_{-k}}$ and $ n^{d}$, $ n^{h}$. Working in the commensurate case, $\rho = \bar n$ and hence $n^{d} = n^{h}$, we solve these equations as detailed in Appendix \ref{sec:strApp}.

We find that the behavior of the system is characterized by two processes with two corresponding time scales. 

The first process, occurring at a rate $1/\tau_{m} \sim \gamma$, involves the localization of quasiparticles when they are detected. It is illustrated by the occupation number of doublons with momentum $k$,
\begin{equation} \begin{split}
& \avg{\hat d_{k}\dg \hat d_{k}}  =  
\br{\avg{\hat d_{k}\dg \hat d_{k}}_{t=0} - n^{d}_{k}}e^{-\gamma t} + n^{d}_{k}
\\ & - e^{-\gamma t}  \Delta_{k} [\tfrac{\gamma^{2}}{U^{2}} \p{1 -\cos{\scriptstyle\p{Ut}}} + \tfrac{\gamma}{U}\sin{\scriptstyle\p{Ut}}]
+ {O\p{\tfrac{J}{U}}^{3}}.
\label{eq:dndk}
\end{split} \end{equation}
Apart from the structure of the transient oscillatory term, this behavior is similar to the weakly-interacting case in Eq.~\pref{eq:dnkweak}. The momentum distribution of the quasiparticles is driven to one which is slowly varying and nearly uniform,
\begin{equation} \begin{split}
& n^{d}_{k} = n^{d} +  \Delta_{k},
\\  \Delta_{k} & = 
\frac{J^{2}}{U^{2}}\frac{2\tilde n^{2}\p{1 - 3 n^{d}}^{2}U^{2}}{\p{1 - 3 n^{d}}^{2}U^{2}+\gamma^{2}}\p{1 -  n^{d}}\p{\gve_{k}^{2} - 2D}.
\end{split} \end{equation}
As is implicit in the form of $\Delta_{k}$, this represents a competition between the coherent creation of quasiparticles and the measurement-induced destruction of coherences.

In parallel, the measurement process results in a slow increase in the total number of quasiparticles. The rate of this process is characterized by $1/\tau_{p} \sim \tfrac{4D\tilde n^{2}J^{2}}{U^{2}+\gamma^{2}} \gamma$ and it is governed by the nonlinear equation of motion
\begin{equation} \begin{split}
\frac{d}{dt} n^{d}  & =  \frac{J^{2}}{U^{2}}\frac{4D\tilde n^{2}\p{1 -3n^{d}}^{2}U^{2}}{\p{1 - 3n^{d}}^{2}U^{2}+\gamma^{2}}\p{1 - n^{d}}\gamma \times
\\ &   
\bmat{1  - e^{-\gamma t}\p{\cos\oUt + \tfrac{\gamma}{U}\sin\oUt} + O\p{\tfrac{J}{U}}}.
\label{eq:dnd}
\end{split} \end{equation}
One intuition for this growth comes from picturing the Mott insulator state as filled with virtual doublon-holon pairs. Whenever a virtual doublon or holon is imaged, the pair is converted into a real doublon and holon.

For shorter times, $\gamma t \ll \p{\tfrac{J}{U}}^{-2}$, the number of excitations remains small, $n^{d} \ll 1$. Then the right hand side of Eq.~\pref{eq:dnd} may be  integrated,
\begin{equation} \begin{split}
& n^{d}  = n^{d}_{t=0} + 
\\ & \frac{4D\tilde n J^{2}}{U^{2}+\gamma^{2}}
  \br{\gamma t - \tfrac{2\gamma^{2}}{U^{2} + \gamma^{2}}\p{1 - e^{-\gamma t}\Xi\p{t}} + O\p{\tfrac{J}{U},n^{d}}},
\label{eq:earlyt}
\end{split} \end{equation}
where the transient oscillations 
\begin{equation} \begin{split}
\Xi\p{t}  =  \cos\oUt  - \half  \p{\tfrac{U}{\gamma} -\tfrac{\gamma}{U}}\sin\oUt
\end{split} \end{equation}
are followed by  linear growth in the excitation density.

\begin{figure}[htbp] 
   \centering
   \includegraphics[width=\columnwidth]{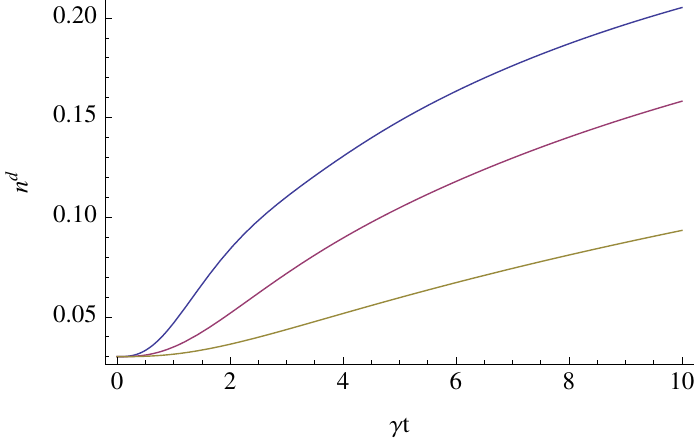} 
   \caption{(Color online) Growth in doublon density with measurement in a Mott system. At short times, the measurement process primarily scatters doublons into a uniform momentum occupation. This is followed by a growth in doublon density that is initially linear and levels off as a result of the hard-core constraints on doublon occupation. From top to bottom (blue, magenta, yellow) $\gamma/U = 0.5,1,2$, in a three-dimensional cubic lattice with $J/U = 0.05 $ , $\bar n = 1$.}
   \label{fig:nd}
\end{figure}

The complete time evolution of $\avg{n^{d}}$ is plotted in Fig.~\ref{fig:nd} for typical parameters. 

Within our approximations, $n^{d}\to \tfrac{1}{3}$ as long times. This is the infinite temperature limit of the model in Eq.~\pref{eq:HDH}: each site is equally likely to be empty, have a doublon or have a holon. However, once $n^{d}$ is of order unity, the model no longer fully describes the physics, and one must include larger fluctuation in the site occupation to fully capture the physics.

\begin{figure}[htbp] 
   \centering
   \includegraphics[width=\columnwidth]{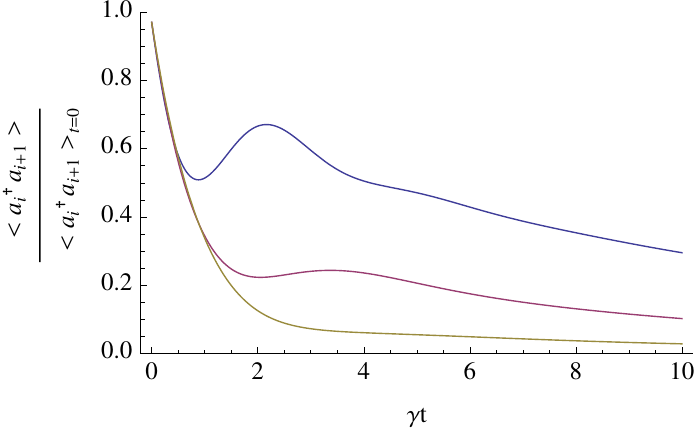} 
   \caption{(Color Online) The evolution of the nearest-neighbor single-particle correlation function $\avg{\hat a_{i}\dg\hat a_{i+1}}$ in a Mott system. From top to bottom (blue, magenta, yellow) $\gamma = 0.5,1,2$, in a three-dimensional cubic lattice with $J = 0.05 U$ , $\bar n = 1$.}
   \label{fig:MottCoh}
\end{figure}

The atom correlation functions can be calculated from those of the doublons and holons. They will be short ranged, dominated by nearest neighbor correlations, such as
\begin{equation} \begin{split}
& \avg{\hat a_{i}\dg \hat a_{i+1}} =
\\ & \quad \frac{J}{U}\frac{2\tilde n^{2}\p{1 - 3n^{d}}^{2}U^{2}}{\p{1 - 3n^{d}}^{2}U^{2}+\gamma^{2}}\p{1 - n^{d}}\times
\\ & \quad \br{1 + e^{-\gamma t}(\tfrac{\gamma^{2}}{U^{2}}\cos\oUt - \tfrac{\gamma}{U}\sin\oUt)+ O\p{\tfrac{J}{U}}}
\end{split} \end{equation}
These are plotted in Fig.~\ref{fig:MottCoh} for typical parameters. As discussed above, two time scale are apparent in the graph.

\section{Long wavelength measurements\label{sec:lowres}}

We have explored so far the destruction of non-local correlations from the spontaneous localization of atoms to single lattice sites. As previously noted \cite{Holland1996,Pichler2010}, the length scale of the localization is determined by the wavelength of the emitted light. Here we extend our argument to the case where the wavelength of light measuring the system is larger than the lattice spacing. 

A simple model of such a measurement is
\begin{equation} \begin{split}
\hat M_{i} = \hat n^{\xi}_{i} = \frac{1}{\mathcal N_{\xi}}\sum_{j}e^{-\half\p{\frac{r_{j}-r_{i}}{\xi}}^{2}}\hat n_{j},
\end{split} \end{equation}
where the normalization is $\mathcal N_{\xi} = \sum_{i} e^{-\half \p{r_{i}/\xi}^{2}}$ is proportional to the width of the measurement. 

Measurement with such long wavelength light does not localize the atoms to single lattice sites. One learns less about the system, but perturbs it proportionally less.

For free particles, the evolution of momentum states is replaced by the equation
\begin{equation} \begin{split}
& \frac{d}{dt}\avg{\hat n_{k}} = -\frac{\mathcal N_{\xi/\sqrt{2}}}{\p{\mathcal N_{\xi}}^{2}}\gamma\p{\avg{\hat n_{k}} - \tfrac{1}{N_{s}}\sum G_{\xi}\p{p} \avg{\hat n_{k+p}}}
\end{split} \end{equation}
where $G_{\xi}\p{p} = \sum_{i} e^{-\tfrac{1}{4}\p{\tfrac{r_{i}}{\xi}}^{2}}e^{ip r_{i}}$.  For the two-point correlation we find the closed form
\begin{equation} \begin{split}
& \qquad \avg{\hat a_{i}\dg \hat a_{j}}  = \avg{\hat a_{i}\dg \hat a_{j}}_{t=0} e^{-\bar\gamma_{\abs{i-j}}t}
\\ & \bar\gamma_{\abs{i-j}}  = g_{\xi} \sqrt{\tfrac{1}{\pi}}\tfrac{\Delta r}{2D\xi}\p{1 - \exp\br{-\tfrac{1}{4}\p{\tfrac{r_{i}-r_{j}}{\xi}}^{2}}},
\end{split} \end{equation}
where $\Delta r$ is the spacing between sites and the function $g_{\xi} = \tfrac{\mathcal N_{\xi/\sqrt{2}}}{\p{\mathcal N_{\xi}}^{2}} \br{\sqrt{\tfrac{1}{\pi}}\tfrac{\Delta r}{2D\xi}}^{-1}$ has $g_{\xi}\approx 1$ for $\abs{\xi}\gtrsim \abs{\Delta r}$. Thus the rate at which correlations are lost is suppressed linearly in $\xi$ and correlations on scales smaller than $\xi$ decay at a much reduced rate.

\section{Summary}

We would like to have non-destructive site-resolved measurements. Unfortunately, no measurement is entirely non-destructive. Here we have quantified the effect of an ideal density measurement on a lattice system. In the superfluid regime, we use Bogoliubov theory to show that all spatial correlations decay exponentially with $\gamma t$, the number of photons scattered. In the Mott regime, we find that the momenta of the quasiparticles are quickly scrambled, leading to a slowly evolving quasi-steady state. In this slow-proliferation stage, fluctuations in the on-site density gradually grow. Similar physics was seen in numerical studies \cite{Poletti2012,Poletti2013}.

We predict how momentum occupation and single-particle correlations evolve with time. The former can be studied through time of flight experiments \cite{Condensation1995}. Protocols exist for the direct measurement of the single particle correlation function \cite{Hadzibabic2006,Richard2003,Stenger1999,Hagley1999}. Finally, though our focus is on measurement, the formalism and all of our results apply to spontaneous emission (in the absence of excitations to higher bands). As such they provide a quantitative estimate of the effects of spontaneous emission on coherence.

It is useful to put the loss of correlations into the context of the information gained as light is emitted.  Assuming no dynamics, the continual measurement reduces the uncertainty in the number of atoms on a given site with time, $\gd n_{i}^{2} \sim e^{-\gamma t}$ \cite{Gregory2005}. Thus, in the superfluid regime, the uncertainty falls at the same rate as do the correlations. In the Mott regime the uncertainty falls faster. 

In this regard, long-wavelength measurements may be advantageous. If one wishes to measure the total number of particles in the cloud, the reduced uncertainty is set by the number of scattered photons, not their wavelength. As seen above, however, the backaction is reduced for long-wavelength probes. In general, one would wish to tailor the process of measurement so that all information carried by the probe is experimentally accessible.

\section*{Acknowledgements}
We thank Mukund Vengalattore and his students for extensive discussions. We acknowledge support from the ARO-MURI Non-equilibrium Many-body Dynamics grant (63834-PH-MUR).

\appendix
\widetext
\section{Derivation of Formulas in the Strongly Interacting Case \label{sec:strApp}}
We present here the full derivation of our results for strongly interacting bosons. 

Our starting point is the Hamiltonian
\begin{equation} \begin{split}
 \hat H_{\rm DH}  & =  -J\sum_{\avg{i,j}}
\br{\bar n\p{\bar n+1}\hat d_{i}\dg\hat d_{j} + \bar n \hat h_{i}\dg\hat h_{j} + \sqrt{\bar n \p{\bar n + 1}}\p{\hat d_{i}\dg\hat h_{j} + \hat h_{i}\dg \hat d_{j}} + \hc}
 + U \sum_{i}\p{\hat d_{i}\dg\hat d_{i} + \hat h_{i}\dg \hat h_{i}}
\\ & =   \sum_{k}\br{\tfrac{U}{2} + J\p{\sqrt{\tilde n^{2} + \tfrac{1}{4}} + \half}\gve_{k}}\hat d_{k}\dg\hat d_{k}
 + \br{\tfrac{U}{2} + J \p{\sqrt{\tilde n^{2} + \tfrac{1}{4}} - \half} \gve_{k}}\hat h_{k}\dg \hat h_{k} + J \tilde n\gve_{k}\p{\hat d_{k}\hat h_{-k} + \hat h_{-k}\dg \hat d_{k}\dg}
\label{eq:HDHApx}
\end{split} \end{equation}
where as before, $\gve_{k} = -2\sum_{\Delta \vec r}\cos\p{\vk\cdot\Delta\vec r}$ and $\tilde n = \sqrt{\bar n \p{\bar n + 1}}$. With this Hamiltonian the difference between the total number of doublons and holons is constant. We work in the commensurate case, where the particle density is given by the integer $\bar n$ and the total number of doublons equals the total number of holons.

The operators $\hat d_{i}$ and $\hat h_{i}$  have a hard core constraint $\hat d_{i}^{2} = \hat h_{i}^{2} = \hat d_{i}\hat h_{i} = 0$. In equilibrium, at small $J/U$ and $T/U$, this constraint has little effect as the densities of doublons and holons is small. During the measurement process, however, the number of quasiparticles grows, and we will need to include these constraints.

\subsection{Initial State}
The initial equilibrium properties of Eq.~\pref{eq:HDHApx} can be calculated by performing a Bogoliubov transformation, 
\begin{equation} \begin{split}
\begin{array}{c}
\hat d_{k} = \cosh\theta_{k}\tilde d_{k} + \sinh\theta_{k}\tilde h_{-k}\dg, 
\\ \hat h_{k} = \cosh\theta_{k}\tilde h_{k} + \sinh\theta_{k}\tilde d_{-k}\dg,
\end{array}
\qquad  \tanh\p{2\theta_{k}}  = -\frac{2J\tilde n\gve_{k}}{U + 2J\sqrt{\bar n^{2} + \tfrac{1}{4}}\gve_{k}}.
\end{split} \end{equation}
Neglecting the hard-core constraints, which can be ignored for low-defect densities, $\tilde d_{k}$, $\tilde h_{k}$ are bosonic operators and the Hamiltonian takes the diagonal form
\begin{equation} \begin{split}
 \hat H_{\rm DHB} & = \sum_{k} \p{\tilde E_{k} + \half J\gve_{k}}\tilde d_{k}\dg\tilde d_{k}  + \p{ \tilde E_{k} - \half J\gve_{k}}\tilde h_{k}\dg\tilde h_{k},
\\ \tilde E_{k} & =  \half\sqrt{\p{U + 2J\sqrt{\tilde n^{2} + \tfrac{1}{4}} \gve_{k}}^{2} - \p{2J\tilde n\gve_{k}}^{2}}.
\label{eq:HDHB}
\end{split} \end{equation}

We take our initial conditions to correspond to the ground state, where $\avg{\tilde d_{k}\dg\tilde d_{k}} = \avg{\tilde h_{k}\dg\tilde h_{k}} = 0$, and hence
\begin{equation} \begin{split}
\avg{\hat d_{k}\dg\hat d_{k}}_{t=0} = \avg{\hat h_{k}\dg\hat h_{-k}}_{t=0} & = \p{\tfrac{J}{U}}^{2}\tilde n^{2}\gve_{k}^{2} + O\p{\tfrac{J}{U}}^{3},
\\ \avg{\hat d_{k}\hat h_{-k}}_{t=0}   =  -\tfrac{J}{U}&\tilde n\gve_{k} + O\p{\tfrac{J}{U}}^{2}.
\label{eq:init}
\end{split} \end{equation}
The calculation may be easily extended to low finite temperatures as long as the initial particle densities remain of the order $\p{\tfrac{J}{U}}^{2}$.

\subsection{Evolution Equations}
To obtain the full evolution equations we must now include the hard core constraints. In momentum space, these constraints lead to the commutation relations
\begin{equation} \begin{split}
\br{\hat d_{k}, \hat d_{q}\dg} & = \gd_{k,q} - \tfrac{1}{N_{s}}\sum_{p}2\hat d_{q+p}\dg\hat d_{k+p} + \hat h_{q+p}\dg\hat h_{k+p}
\\ \br{\hat h_{k}, \hat h_{q}\dg} & = \gd_{k,q} - \tfrac{1}{N_{s}}\sum_{p}\hat d_{q+p}\dg\hat d_{k+p} + 2\hat h_{q+p}\dg\hat h_{k+p}
\\ &\br{\hat d_{k}, \hat h_{q}\dg}  = - \tfrac{1}{N_{s}}\sum_{p}\hat h_{q+p}\dg\hat d_{k+p}.
\label{eq:com}
\end{split} \end{equation}
In these sums, the terms where operators have different momentum indices will add incoherently, suggesting the approximation
\begin{equation} \begin{split}
\br{\hat d_{k}, \hat d_{q}\dg} & \approx \gd_{k,q}\p{1 - 2{\hat n^{d}} - {\hat n^{h}}},
\qquad \br{\hat h_{k}, \hat h_{q}\dg}  \approx \gd_{k,q}\p{1 - 2{\hat n^{h}} - {\hat n^{d}}}, \qquad \br{\hat d_{k}, \hat h_{q}\dg} \approx 0,
\label{eq:Apcom}
\end{split} \end{equation}
where $\hat n^{d} = \tfrac{1}{N_{s}}\sum_{k}\hat d_{k}\dg\hat d_{k}$ and similarly for $\hat n^{h}$.

We substitute Eq.~\pref{eq:HDHApx} into Eq.~\pref{eq:dObs}, using the commutators in Eq.~\pref{eq:Apcom}, for $\hat O = \hat d_{k}\dg \hat d_{k}$ and  $\hat O = {\hat d_{k} \hat h_{-k}}$. We assume that the total number of quasiparticles is uncorrelated with their momentum distribution,
\begin{equation} \begin{split}
\avg{\hat n^{d}\hat d_{k}\dg\hat d_{k}} & \approx n^{d}\avg{\hat d_{k}\dg\hat d_{k}}, 
\qquad \avg{\hat n^{d}\hat h_{k}\dg\hat h_{k}}  \approx n^{d}\avg{\hat h_{k}\dg\hat h_{k}},
\qquad \avg{\hat n^{d}\hat d_{k}\hat h_{-k}}  \approx n^{d}\avg{\hat d_{k}\hat h_{-k}},
\\ \avg{\hat n^{h}\hat d_{k}\dg\hat d_{k}} & \approx  n^{h}\avg{\hat d_{k}\dg\hat d_{k}}, 
\qquad \avg{\hat n^{h}\hat h_{k}\dg\hat h_{k}}  \approx n^{h}\avg{\hat h_{k}\dg\hat h_{k}},
\qquad \avg{\hat n^{h}\hat d_{k}\hat h_{-k}}  \approx n^{h}\avg{\hat d_{k}\hat h_{-k}},
\end{split} \end{equation}
where $n^{d,h} = \avg{\hat n^{d,h}}$. One can formally derive these relations through perturbation theory in $J/U$, although their validity is wider.

The evolution equations then simplify to a set of coupled nonlinear differential equations,
\begin{equation} \begin{split}
\frac{d}{dt} \avg{\hat d_{k}\dg \hat d_{k}} & =  -2\tilde n \bar J_{t}\gve_{k}\Im\br{\avg{\hat d_{k}\hat h_{-k}}} 
- \gamma\p{\avg{\hat d_{k}\dg \hat d_{k}} - n^{d}}
\\ \frac{d}{dt} \avg{\hat d_{k} \hat h_{-k}} & =  -i\tilde n\bar J_{t}\gve_{k} P_{t} -i\p{\bar U_{t} + 2\sqrt{\tilde n^{2} + \tfrac{1}{4}} \bar J_{t}\gve_{k} }  \avg{\hat d_{k}\hat h_{-k}} 
- \gamma \avg{\hat d_{k}\hat h_{-k}}.
\label{eq:timeevol}
\end{split} \end{equation}
where all $k$ dependence is through $\gve_{k} = -2\sum_{\Delta \vec r}\cos\p{\vk\cdot\Delta\vec r}$. Here
\begin{equation} \begin{split}
\bar J_{t} = J\p{1 - 3n^{d}}, & \qquad \bar U_{t}  = U \p{1 - 3n^{d}}, \qquad P_{t} = \p{1 + 2\avg{\hat d_{k}\dg \hat d_{k}} - 3n^{d}}
\end{split} \end{equation}
are time dependent, but, as we will see, vary at a rate much slower than $\gamma$.

In the commensurate case, $n^{d} =  n^{h}$, and one finds identical initial values and evolution equations for the momentum occupation of holons and doublons, hence $\avg{\hat d_{k}\dg \hat d_{k}} = \avg{\hat h_{k}\dg \hat h_{k}}$ at all times.

\subsection{Ansatz Solution}
All of the $k$-dependence in Eq.~\pref{eq:timeevol} arises from terms of the form $J\gve_{k}$. Since $J\ll U$, we can expand in this product, finding
\begin{equation} \begin{split}
\avg{\hat d_{k}\dg \hat d_{k}} = \avg{\hat h_{-k}\dg \hat h_{-k}} & = dd^{\p{0}} + dd^{\p{2}}\p{\tfrac{J}{U}}^{2}\gve_{k}^{2} + O\p{\tfrac{J}{U}}^{3}
\\ \avg{\hat d_{k} \hat h_{-k}} & =  dh^{\p{1}}\p{\tfrac{J}{U}}\gve_{k} + O\p{\tfrac{J}{U}}^{2}.
\end{split} \end{equation}
where $ dd^{\p{0}}, dd^{\p{2}}, dh^{\p{1}}$ are functions of time but not $k$. By Fourier transforming these expressions we can relate them to the more familiar
\begin{equation} \begin{split}
n^{d} = dd^{\p{0}} + 2D\p{\tfrac{J}{U}}^{2} dd^{\p{2}} + O\p{\tfrac{J}{U}}^{3}, & \qquad \avg{\hat d_{i}\hat h_{i+1}} = -\tfrac{J}{U}dh^{\p{1}} + O\p{\tfrac{J}{U}}^{2}.
\end{split} \end{equation}

Eqs.~\pref{eq:timeevol} then reduce to
\begin{eqnarray}
\frac{d}{dt} n^{d}  & = &  \p{\tfrac{J}{U}}4D \tilde n \bar J_{t}\Im\br{ \avg{\hat d_{i} \hat h_{i+1}}/\tfrac{J}{U}} + O\p{\tfrac{J^{3}}{U^{2}}}
\label{eq:dndAns}
\\ \frac{d}{dt} \avg{\hat d_{i} \hat h_{i+1}} & =  &
 i\tilde n\bar J_{t}\p{1  - n^{d}} -i\bar U_{t} \avg{\hat d_{i} \hat h_{i+1}}  - \gamma \avg{\hat d_{i} \hat h_{i+1}} + O\p{\tfrac{J^{2}}{U}}
 \label{eq:ddhAns}
\\\frac{d}{dt} dd^{\p{2}} & = & 2\tilde n\bar U_{t}  \Im\br{ \avg{\hat d_{i} \hat h_{i+1}}/\tfrac{J}{U}} - \gamma dd^{\p{2}}+ O\p{J}
 \label{eq:ddd2Ans}
\end{eqnarray}
while the initial conditions are
\begin{equation} \begin{split}
n^{d}_{t=0} =  \p{\tfrac{J}{U}}^{2}2D&\tilde n^{2}, \qquad dd^{\p{2}}_{t=0} = \tilde n^{2},
\qquad \avg{\hat d_{i} \hat h_{i+1}}_{t=0} = \tfrac{J}{U}\tilde n.
\label{eq:dAnsInit}
\end{split} \end{equation}

We note that Eq,~\pref{eq:dndAns} and \pref{eq:ddhAns} are coupled to each other but independent of Eq.~\pref{eq:ddd2Ans}. At this point, the equations may be numerically integrated for any given values of $\gamma, J, U$. Typical values are plotted in in Figs.~\ref{fig:nd}, \ref{fig:MottCoh}.

\subsection{Short Time Behavior}
The initial and short-time behavior of Eqs.~\pref{eq:dndAns}-\pref{eq:ddhAns} can be analyzed using $n^{d}_{t=0} \sim \p{\tfrac{J}{U}}^{2}$. Thus, we can neglect the non-linear terms, $\bar J_{t} \approx J, \bar U_{t} \approx U$ finding
\begin{equation} \begin{split}
\frac{d}{dt} n^{d}  & =  \tfrac{J}{U}4D \tilde n J\,\Im\br{ \avg{\hat d_{i} \hat h_{i+1}}/\tfrac{J}{U}} + O\p{\tfrac{J^{3}}{U^{2}},n^{d}}
\\ \frac{d}{dt} \avg{\hat d_{i} \hat h_{i+1}} & =  i \tilde n J -iU \avg{\hat d_{i} \hat h_{i+1}}  - \gamma \avg{\hat d_{i} \hat h_{i+1}} + O\p{\tfrac{J^{2}}{U},n^{d}}.
\label{eq:dshortt}
\end{split} \end{equation}
The second equation produces a function which oscillates with frequency $U$ while decaying at a rate $\gamma$ to a steady state value,
\begin{equation} \begin{split}
\avg{\hat d_{i} \hat h_{i+1}} = \br{\avg{\hat d_{i} \hat h_{i+1}}_{t=0} - \tfrac{iJ\tilde n}{iU + \gamma}}e^{-\gamma t}e^{-iU t}
+ \tfrac{iJ\tilde n}{iU + \gamma}.
\label{eq:dhshortt}
\end{split} \end{equation}
Using this result to calculate the number of doublons, we find
\begin{equation} \begin{split}
n^{d} = n^{d}_{t=0} + \frac{4D \tilde n^{2}J^{2}}{U^{2} + \gamma^{2}}\br{
\gamma t - \tfrac{2\gamma^{2}}{U^{2}+\gamma^{2}}\p{1 - e^{-\gamma t}\br{\cos\p{U t} - \half\p{\tfrac{U}{\gamma} - \tfrac{\gamma}{U}}\sin\p{U t}} }}.
\label{eq:ndshortt}
\end{split} \end{equation}

Aside from small transients, we see a linear increase in $n^{d}$ with characteristic rate $1/\tau_{p} = \frac{4D \tilde n^{2}J^{2}}{U^{2} + \gamma^{2}}\gamma$. Physically, this is the rate at which virtual doublon-holon pairs are imaged. This linearized theory breaks down when $n^{d}\sim 1$. Thus it is valid until $ t\sim \tau_{p}\gg 1/\gamma$.

\subsection{General Behavior}
Given the separation of timescales between the rate of change in $n^{d}$ and $\avg{\hat d_{i}\hat h_{i+1}}$, we can adiabatically eliminate the nonlinear terms in Eq.~\pref{eq:dndAns}, rather than simply neglecting them. This yields
\begin{equation} \begin{split}
\avg{\hat d_{i} \hat h_{i+1}}  = & \p{\avg{\hat d_{i} \hat h_{i+1}}_{t=0} - \avg{\hat d_{i} \hat h_{i+1}}^{long}} e^{-\gamma t}e^{-iU t} + \avg{\hat d_{i} \hat h_{i+1}}^{long}
+ O\p{\tfrac{J}{U}}^{2}
\\ &\avg{\hat d_{i} \hat h_{i+1}}^{long} =  \frac{i\tilde n J\p{1 - 3n^{d}}}{iU\p{1 - 3n^{d}} + \gamma} \p{1  - n^{d}}
\end{split} \end{equation}
at all times. When $n^{d} \ll 1$, this reduces to Eq.~\pref{eq:dhshortt}.

Substituting this into Eq.~\pref{eq:dndAns} yields
\begin{equation} \begin{split}
\frac{d}{dt} n^{d} = \frac{J^{2}}{U^{2}}\frac{4D \tilde n^{2}\p{1 - 3n^{d}}^{2}U^{2}}{\p{1 - 3n^{d}}^{2}U^{2} + \gamma^{2}}\p{1 - n^{d}}\gamma\br{
1 - e^{-\gamma t}\p{\cos\p{U t} + \tfrac{\gamma}{U}\sin\p{U t}}}
\end{split} \end{equation}
which simplifies to Eq.~\pref{eq:dshortt} for $n^{d} \ll 1$. Likewise, we adiabatically eliminate Eq.~\pref{eq:ddd2Ans} to obtain Eq.~\pref{eq:earlyt} in the main text.

\bibliographystyle{apsrev}
\bibliography{/Users/yarivyanay/Documents/University/Citations/library}

\end{document}